\documentclass{article}
\usepackage{spconf,amsmath,graphicx}
\usepackage{url}
\usepackage{amssymb}
\usepackage{pifont}
\usepackage{hyperref}

\title{EV-NVC: efficient variable bitrate neural video compression}
\name{Yongcun Hu\textsuperscript{+}, Yingzhen Zhai\textsuperscript{+}, Jixiang Luo\textsuperscript{*}, Wenrui Dai, Dell Zhang, Hongkai Xiong, Xuelong Li\textsuperscript{*} }
\address{ 
huyongcun61@gmail.com, zhaiyingzhen23@mails.ucas.ac.cn \\
luojx14@chinatelecom.cn, daiwenrui@sjtu.edu.cn, \\ 
dell.z@ieee.org, xionghongkai@sjtu.edu.cn, xuelong\_li@ieee.org
 \\
Institute of Artificial Intelligence (TeleAI), China Telecom}

\begin{document}

%
\maketitle
\footnotetext[1]{+ Intern at TeleAI and equal contribution}
\footnotetext[2]{* Corresponding author}

\begin{abstract}
Training neural video codec (NVC) with variable rate is a highly challenging task due to its complex training strategies and model structure. In this paper, we train an efficient variable bitrate neural video codec (EV-NVC) with the piecewise linear sampler (PLS) to improve the rate-distortion performance in high bitrate range, and the long-short-term feature fusion module (LSTFFM) to enhance the context modeling. Besides, we introduce mixed-precision training  and discuss the different training strategies for each stage in detail to fully evaluate its effectiveness. 
Experimental results show that our approach reduces the BD-rate by 30.56\% compared to HM-16.25 within low-delay mode. 

\end{abstract}
\section{Introduction and related work}
\label{sec:intro}


With the rapid development of short video platforms and self-media, there has been an explosive increase in the number of videos available on the internet. 
While over the past two decades, traditional video codecs have undergone significant development to meet evolving user demands, progressing from H.264/AVC \cite{wiegand2003overview} to H.265/HEVC \cite{sullivan2012overview}, and then to H.266/VVC \cite{bross2021overview}. 

In recent years, convolutional neural network (CNN)-based end-to-end video compression algorithms have witnessed substantial advancements, achieving superior performance compared to traditional methods. These innovations can be traced back to the early development of end-to-end image compression algorithms \cite{balle2016end, balle2018variational, luo2024rethinking}, from which they have naturally evolved into the domain of video processing. Such algorithms can generally be classified into three main types: 3D autoencoder-based algorithms, residual coding-based algorithms, and conditional coding-based algorithms. The first category treats videos as a temporal sequence of images and extends the conventional 2D autoencoder used in image compression to a 3D autoencoder \cite{habibian2019video, pessoa2020end, sun2020high}, achieving video compression. Although this approach is relatively straightforward to implement, it typically demands large amounts of memory and computational resources. Residual coding-based algorithms \cite{hu2021fvc, agustsson2020scale, shi2022alphavc} replace traditional codec components with neural networks. These methods utilize an optical flow extractor to estimate the optical flow between adjacent frames, which, along with the decoded previous frame, is used for warping to predict the current frame. The predicted frame is then combined with the encoded and transmitted residual to reconstruct the final frame. Rather than residual coding-based algorithms, conditional coding-based algorithms \cite{li2021deep, li2022hybrid, li2023neural, li2024neural} treat the decoded previous frame as condition information and calculate the reconstructed current frame implicitly using a conditional neural network. Thus, this approach is capable of achieving a lower entropy bound \cite{ladune2020optical}. 

Conditional coding-based algorithms originate from \cite{li2021deep} and have gained great attention in recent years. \cite{sheng2022temporal} not only preserves the reconstructed frames,  but also retains intermediate features, which are used as additional information for encoding and decoding the subsequent frame. With incorporating multi-scale feature fusion during motion compensation, the method achieves performance comparable to VTM in low-delay mode. \cite{li2023neural} introduces a group-based offset diversity technique to capture optical flow, while also incorporating a quadtree-based partitioning scheme within the entropy model. This combination effectively balances computational efficiency with accuracy, achieving performance comparable to that of ECM. \cite{li2024neural} propose a variable rate strategy and a refresh mechanism, making it a variable rate SOTA model in low-latency coding.

Although neural video codecs have demonstrated high performance, several critical issues remain unresolved. Firstly, the majority of existing algorithms focus primarily on reconstructing the current frame based on either the previous frame or latent features, neglecting the integration of information from earlier frames or features. This limitation results in suboptimal reconstruction quality. Secondly, many of the existing models do not support variable rate, necessitating multiple rounds of training and storage during operation, which leads to inefficient use of resources. Thirdly, while the aforementioned models typically require multi-stage training, there is a notable lack of comprehensive analysis regarding the specific training strategies applied at each stage. These absences significantly hamper the rapid reproduction and validation of these algorithms.

In this paper, we report EV-NVC, an open-source NVC integrating with a varible rate strategy, a long- and short-term fusion strategy, and a multi-stage training strategy. With these components, our EV-NVC outperforms VTM-17.0 in the YUV420 colorspace. Specifically, our contributions are:
\begin{itemize}
    \item We propose a piecewise linear sampler (PLS) for training variable rate models, which effectively improves the rate-distortion performance in high bitrate range.
    \item We propose a long-short-term feature fusion module (LSTFFM), which reduces the BD-Rate by further fusing long-term information.
    \item We propose a multi-stage and mixed-precision training strategy and offer a compressive analysis of the objective functions and key parameters employed at each stage.

\end{itemize}

\section{overview of our approach}
\label{sec:format}
\begin{figure}
    \centering
    \includegraphics[width=0.99\linewidth]{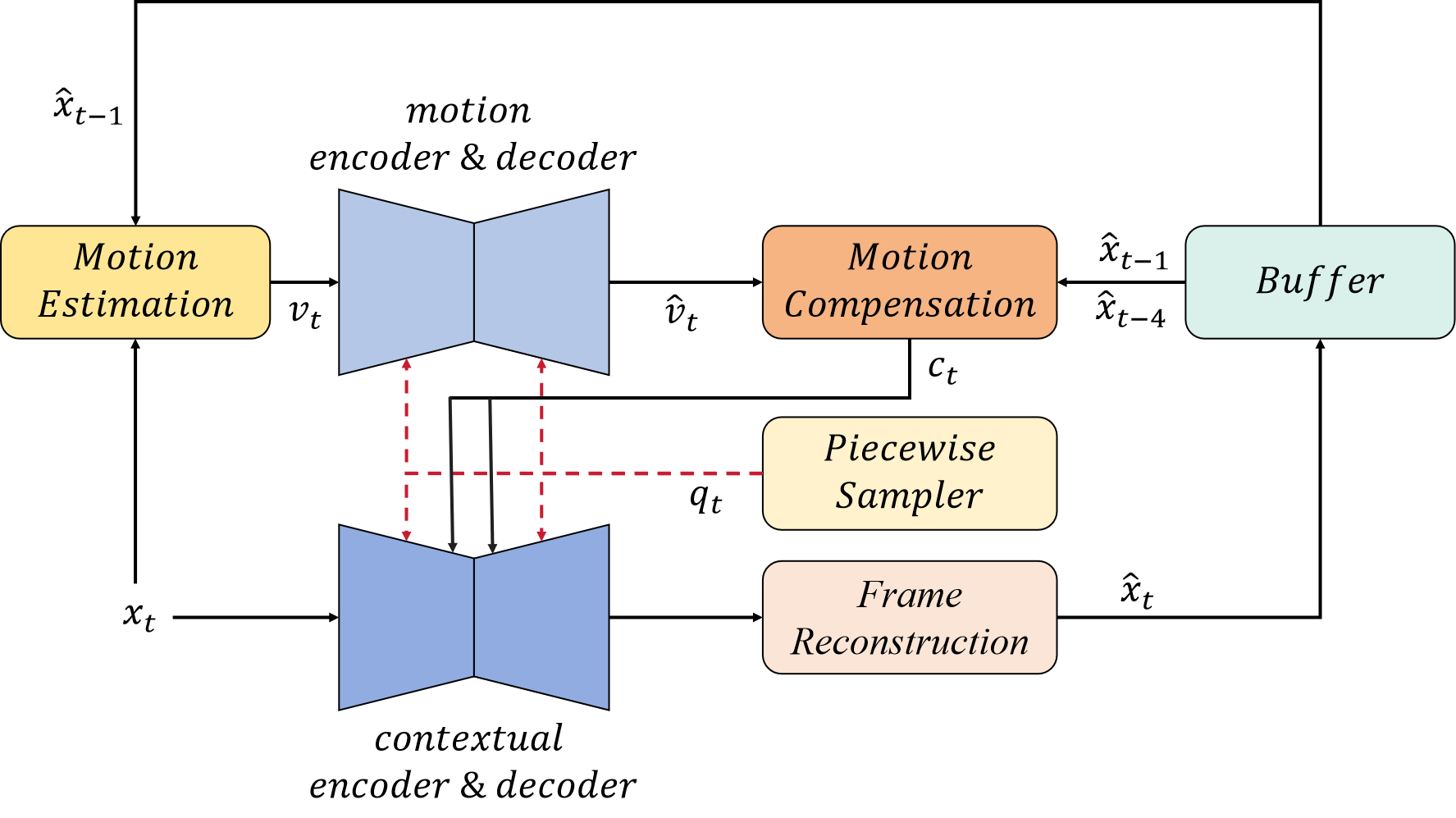} 
    \caption{Overview of the proposed method. $x_{t}$ and $\hat{x}_{t}$ are current and reconstructed frame. $v_{t}$ and $\hat{v}_{t}$ are motion vector and decoded motion vector. $c_{t}$ is the conditional information. $q_{t}$ is the rate control parameter.}
    \label{fig:overview}
\end{figure}

Fig.~\ref{fig:overview} illustrates the framework of the proposed method. For clarity, the entropy model is omitted. During training, we introduce PLS to generate the rate control parameter $q$. Compared to uniform sampling, this approach improves performance in the high bitrate range. The design of the sampler is discussed in Section~\ref{ssec:varible bitrate piecewise sampling strategy}. In terms of \textit{motion compensation}, we employ LSTFFM to further integrate long-term information, as detailed in Section~\ref{ssec:long- and short-term fusion strategy}. For \textit{motion estimation}, we adopt the pre-trained SpyNet \cite{ranjan2017optical}. The remaining components of the framework are consistent with DCVC-FM \cite{li2024neural}. 

\section{the proposed approach}
\label{sec:pagestyle}

\subsection{Piecewise Linear Sampler}
\label{ssec:varible bitrate piecewise sampling strategy}

Previous studies, such as \cite{tong2023qvrf}, have observed performance degradation in the high bitrate range when training variable bitrate models using the same set of Lagrange multipliers adopted in fixed bitrate models. To address this issue, it was proposed that expanding the set of Lagrange multipliers to a larger range could mitigate this performance degradation. Inspired by this observation, we hypothesize that variable bitrate models operating in high bitrate and low bitrate ranges should not be treated equally during training. Models in a higher bitrate range are required to recover finer details, making their training more challenging. Therefore, we propose the following piecewise linear sampler.
\begin{equation}
    \label{equ:piecewise-sampling}
    p(idx)=
        \begin{cases}
            \frac{1}{1+m+m^2+m^3} & idx \in [1, \frac{n}{4}] \\
            \frac{m}{1+m+m^2+m^3} & idx \in [\frac{n}{4}, \frac{n}{2}] \\
            \frac{m^2}{1+m+m^2+m^3} & idx \in [\frac{n}{2}, \frac{3n}{4}] \\
            \frac{m^3}{1+m+m^2+m^3} & idx \in [\frac{3n}{4}, n]
        \end{cases}
\end{equation}
\begin{equation}
    \label{equ:cal_lambda}
    \lambda_{idx} = \lambda_{min} * (\frac{\lambda_{max}}{\lambda_{min}})^{\frac{idx}{n}}
\end{equation}
We sample different \textit{idx} during training based on the probabilities in Eq.~\ref{equ:piecewise-sampling}. The sampled \textit{idx} is used to calculate the rate control parameter $q$. Specifically, we maintain a set of $q_{min}$ and $q_{max}$, and perform exponential interpolation based on \textit{idx} to obtain the current $q$. The Lagrange multiplier is calculated using Eq.~\ref{equ:cal_lambda}, as \cite{li2024neural}. In our implementation, the bitrate range is divided into four segments, with the sampling probability of each segment increasing according to a power function. The hyperparameters are configured as $m=2, n=64, \lambda_{min}=0.002, \lambda_{max}=0.25$ in our experiments.

\subsection{Long-Short-Term Feature Fusion Module}
\label{ssec:long- and short-term fusion strategy}
\begin{figure}
    \centering
    \includegraphics[width=0.99\linewidth]{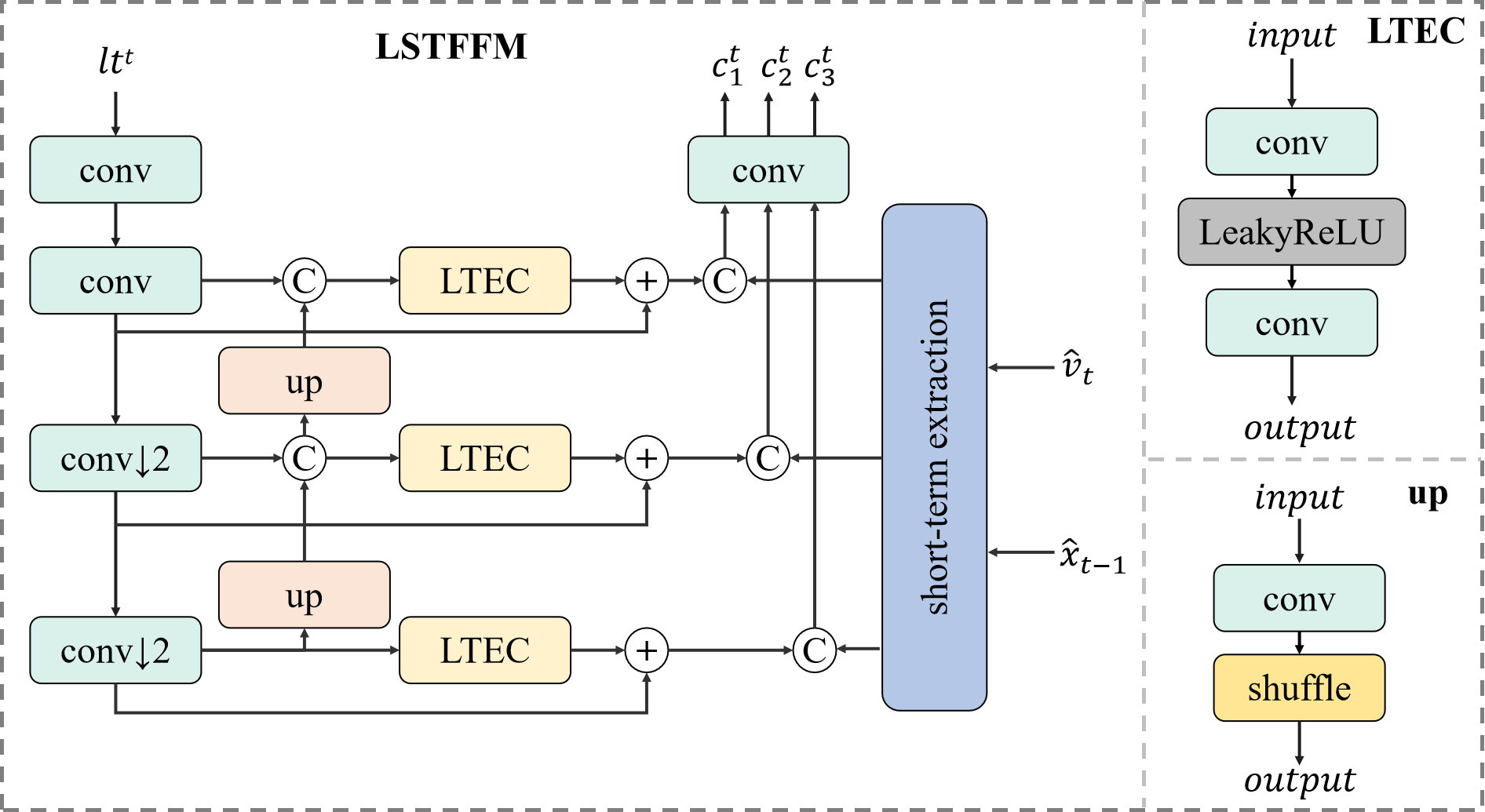} 
    \caption{The structure of long-short-term feature fusion module (LSTFFM). Long-term extraction conv (LTFC) adopts the conv\text{-}LeakyReLU\text{-}conv structure for further feature extraction, and upsampling module is implemented using conv and pixelshuffle. $\mathrm{C}$ is concatentation operation and $\downarrow 2$ is convolutional layer with stride 2.}
    \label{fig:LST module}
\end{figure}


We incorporate long-term information in a simple way without increasing the storage burden. Specifically, when encoding and decoding current frame $x_{t}$, we use not only previous frame $\hat{x}_{t-1}$, but also $\hat{x}_{t-4}$ as reference information. We employ a long-short-term feature fusion module (LSTFFM) to perform feature fusion. 
\begin{equation}
    \label{equ:lt}
    lt^{t} = 
        \begin{cases} \hat{x}_{0} & t < 4 \\
        \hat{x}_{t-4} & \text{else}
        \end{cases}
\end{equation}

Fig.~\ref{fig:LST module} illustrates the structure of the LSTFFM. The right side of the module is the short-term feature extraction component, which corresponds to the multi-scale feature extraction mechanism in DCVC-FM \cite{li2024neural}. This component is specifically designed to capture short-term information. On the left side, the input is processed in accordance with Eq.~\ref{equ:lt}, utilizing a Feature Pyramid Network (FPN) \cite{lin2017feature} with residual connections to perform multi-scale feature extraction for long-term information. The long-term and short-term features are then concatenated and passed through subsequent convolutional layers for feature fusion.

\subsection{Multi-Stage Strategy}
\label{ssec:multi-stage strategy}

\begin{table}
    \centering 
    \caption{Multi-stage strategy.}
    \begin{tabular}{ccccc}
        \hline 
        stage & loss type & frames & lr & segment type \\ \hline 
        stage1 & meD & 2 & 1e-4 & IP \\ 
        stage2 & meRD & 2 & 1e-4 & IP \\ 
        stage3 & recD & 2 & 5e-5 & IP \\ 
        stage4 & meRD & 3 & 1e-4 & PP \\ 
        stage5 & recD & 3 & 5e-5 & PP \\ 
        stage6 & recD & 4 & 5e-5 & PP \\ 
        stage7 & recD & 6 & 5e-5 & PP \\ 
        stage8 & recRD & 2 & 5e-5 & IP \\ 
        stage9 & recRD & 3 & 5e-5 & PP \\ 
        stage10 & recRD & 4 & 5e-5 & PP \\ 
        stage11 & recRD & 6 & 5e-5 & PP \\ 
        stage12 & all & 2 & 5e-5 & IP \\ 
        stage13 & all & 3 & 5e-5 & PP \\ 
        stage14 & all & 4 & 5e-5 & PP \\ 
        stage15 & all & 6 & 5e-5 & PP \\ 
        stage16 & all & 6 & 1e-5 & PP \\ 
        stage17 & all & 6 & 5e-6 & PP \\ 
        stage18 & avg & 6 & 1e-5 & IPP \\ \hline
        \label{tab:table1}
    \end{tabular}
\end{table}

Our training strategy is depicted in Table~\ref{tab:table1} . Following the work from \cite{sheng2022temporal} and \cite{sheng2024nvc}, we adopt varying frame lengths, loss functions, and learning rates at different stages of training. We divide a video into four types: $\textit{I}$, $\textit{IP}$, $\textit{PP}$, and $\textit{IPP}$, and train these objectives sequentially. (1) $\textit{I}$ represents reconstructing I-frame. We use the I-frame model from \cite{li2024neural}. Therefore, we do not train with single frame. (2) $\textit{IP}$ refers to reconstructing P-frames with reference to the I-frame. Therefore, we use two-frame input to train for this objective. (3) $\textit{PP}$ refers to reconstructing P-frames with reference to the previous P-frame. For this objective, a three-frame input is employed during training. Additionally, to match the compression losses inherent in longer sequences, four-frame and six-frame inputs are also used to train. (4) $\textit{IPP}$ refers to reconstructing the current P-frame with reference to the entire decoded sequence. Unlike the previous three loss types, this objective requires an RNN-based approach for training. 
Therefore, we use the longest possible sequence for training.

The rationale for using different frame lengths during training has been discussed above. In the following, we provide a detailed explanation of the selection of different loss functions. In the first four stages, we focus primarily on training the network parameters related to motion vectors with $loss_{meD}$ and $loss_{meRD}$, including the motion vector encoder, motion vector decoder, and the entropy model of motion vector. We use the decoded motion vectors $\hat{mv}$ and the decoded previous frame $\hat{x}_{t-1}$ to perform a warp operation to obtain the current predicted frame $\hat{x}_{t}$.
\begin{equation}
    \label{equ: x_hat_mv}
    \hat{x}_{t} = Warp(\hat{x}_{t-1}, \hat{mv}) \\
\end{equation}
\begin{equation}
    \label{equ: meD}
    loss_{meD}^{t} = \|{x_{t}-\hat{x}_{t}}\|_2^2 \\
\end{equation}
\begin{equation}
    \label{equ: meRD}
    loss_{meRD}^{t} = \lambda * loss_{meD}^{t}+bpp_{mv}^{t} \\
\end{equation}

From stage5 to stage10, we freeze network 
 parameters related to motion vectors and train the remaining network parameters, further improving the video codec performance. At this time, the reconstructed frame $\hat{x}_{t}$ is the output of the entire P-frame model. 
\begin{equation}
    \label{equ: recD}
    loss_{recD}^{t} = \|{x_{t}-\hat{x}_{t}}\|_2^2 \\
\end{equation}
\begin{equation}
    \label{equ: recRD}
    loss_{recRD}^{t} = \lambda * loss_{recD}^{t}+bpp_{context}^{t} \\
\end{equation}

From stage12 to stage17, we train all network parameters. In stage18, we adopt the multi-frame average loss.
\begin{equation}
    \label{equ: all}
    loss_{all}^{t} = loss_{recRD}^{t}+bpp_{mv}^{t} \\
\end{equation}
\begin{equation}
    \label{equ: avg}
    loss_{avg}^{T} = \frac{1}{T}\sum_{t=1}^{T}loss_{all}^{t} \\
\end{equation}

\begin{figure*}[ht]
\begin{minipage}[ht]{0.24\textwidth}
\includegraphics[width=\linewidth]{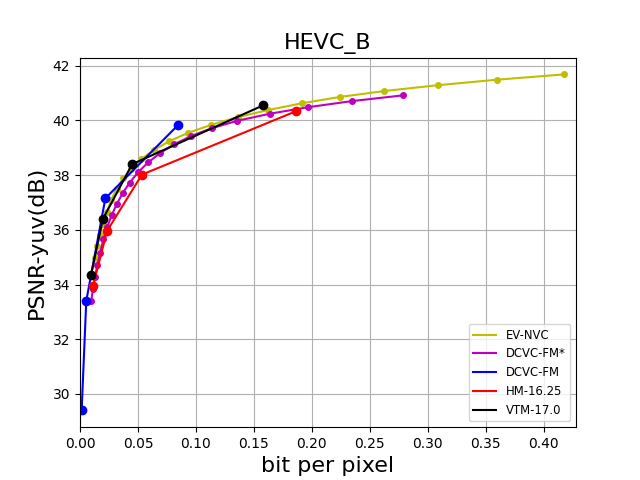}
\end{minipage}
\hfill 
\begin{minipage}[ht]{0.24\textwidth}
\includegraphics[width=\linewidth]{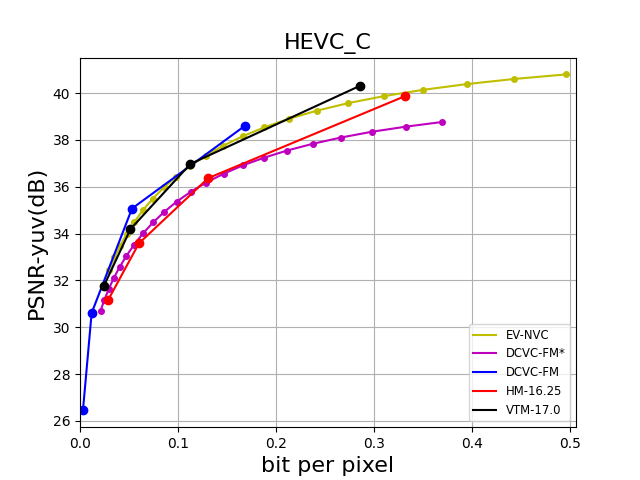}
\end{minipage}
\hfill 
\begin{minipage}[ht]{0.24\textwidth}
\includegraphics[width=\linewidth]{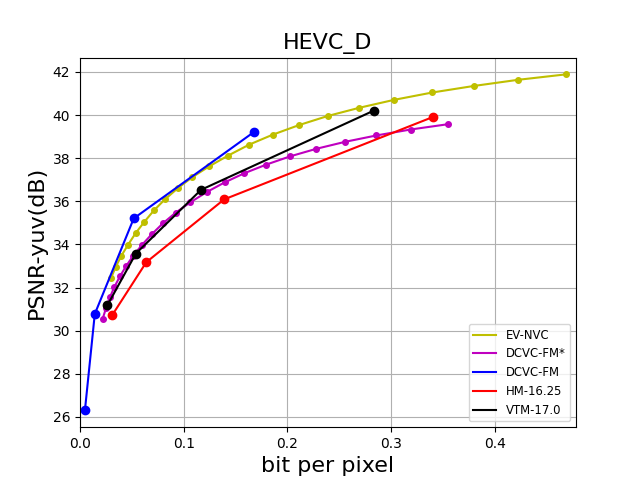}
\end{minipage}
\begin{minipage}[ht]{0.24\textwidth}
\includegraphics[width=\linewidth]{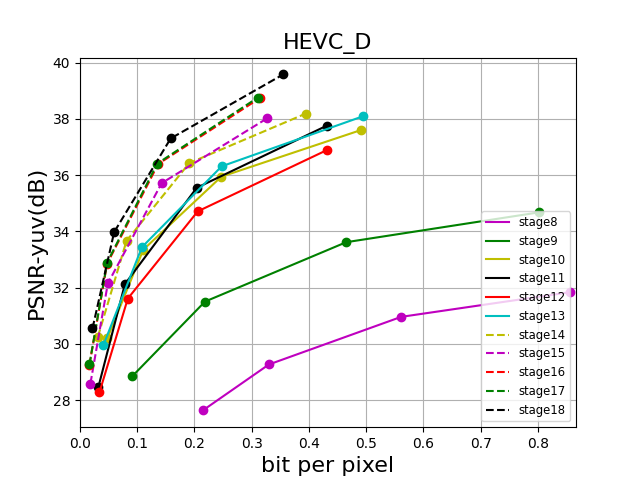}
\end{minipage}

\caption{Rate-distortion curves for HEVC Class B, C, D. The comparison is in YUV420 colorspace. PSNR (6*PSNRY + PSNRU + PSNRV)/8 is used to evaluate the distortion of the decoded pictures. The most right figure is the rate-distortion curve for HEVC Class D from stage8 to stage18.}
\label{fig:Comparisons bdrate}
\end{figure*}

\section{experimental results}
\label{sec:experimental results}
\subsection{Experimental Setup}
\label{ssec:experimental setup}
\textbf{Datasets.} We use Vimeo-90k setuplet dataset \cite{xue2019video} for training, which contains 89800 video sequences and each with 7 frames at a resolution of $448\times256$. The videos are randomly cropped into $256\times256$ patchs. We use RTX 4090 for training, and due to memory limitations, we do not employ additional datasets for fine-tuning. For testing, we use HEVC Class B, C, D \cite{bossen2013common}. 

\textbf{Test condition.} We test 96 frames for each video in YUV420 colorspace. The intra period is set to -1 rather than 32. Our benchmarks include HM-16.25 \cite{HM}, VTM-17.0 \cite{VTM} and DCVC-FM*. DCVC-FM* is the result we train DCVC-FM from scratch, and can be treated a variant of EV-NVC without PLS and LSTFFM, serving as the baseline in the ablation study. 

We use  the Adam \cite{kingma2014adam} optimizer and training each stage for 20 epochs following the Table~\ref{tab:table1}.
To further accelerate training and reduce memory consumption, we employ mixed-precision training. We use BD-Rate \cite{bjontegaard2001calculation} to quantify compression efficiency variations, where negative values correspond to bitrate savings and positive values indicate increased bitrate requirements. 

\subsection{Comparisons With Other Methods}
\label{ssec:Comparisons With Other Methods}

\begin{table}
    \centering 
    \caption{BD-Rate(\%) comparison in YUV420 colorspace measured with PSNR. 96 frames with intra-period=–1. The anchor is HM-16.25.} 
    \begin{tabular}{ccccc}
        \hline 
         & B & C & D & avg \\ \hline 
        HM-16.25 & 0.0 & 0.0 & 0.0 & 0.0 \\ 
        VTM-17.0 & -28.38 & -27.23 & -24.53 & -26.71 \\ 
        DCVC-FM\textsuperscript{*} & -6.70 & -1.76 & -21.86 & -10.11 \\ 
        EV-NVC & -23.49 & -27.41 & -40.79 & -30.56 \\ \hline
    \end{tabular}
    \label{tab:Comparisons bdrate}
\end{table}

Table~\ref{tab:Comparisons bdrate} shows the comparison for YUV420 colorspace under 96 frames with intra-period -1. Our EV-NVC BD-Rate saving over HM and VTM is 30.56\% and 3.85\%, respectively. Despite the removal of PLS and LSTFFM, our DCVC-FM* still achieves a 10.11\% reduction in BD-Rate compared to HM, due to the proposed training strategy. 

Fig.~\ref{fig:Comparisons bdrate} shows the rate-distortion curves. The proposed EV-NVC is capable of selecting a total of 64 RD points and we select one point every four points to plot the curves. The $idx$ of RD points in DCVC-FM \cite{li2024neural} are \{0, 21, 42, 63\}. 
As observed in the figure, EV-NVC covers a high bitrate range and fully encompasses the bitrate range used by HM and VTM. In addition, it exhibits a performance curve that closely aligns with that of VTM on Class B and C, while exhibiting a significant improvement over VTM on Class D. Furthermore, the performance curve of the proposed method is highly comparable to that of DCVC-FM \cite{li2024neural}, despite utilizing a maximum of six frames for training, whereas the latter incorporates an additional 32 frames for post-finetuning. 


\subsection{Ablation Study}
\label{ssec:Ablation Study}

\begin{table}
    \centering 
    \caption{Ablation study for our method.}
    \label{tab:Ablation}
    \begin{tabular}{cccccc}
        \hline 
        PLS & LSTFFM & B & C & D & avg \\ \hline 
        \ding{55} & \ding{55} & 0.0 & 0.0 & 0.0 & 0.0 \\ 
        \ding{51} & \ding{55} & -4.91 & -9.24 & -9.39 & -7.85 \\ 
        \ding{55} & \ding{51} & -8.52 & -10.81 & -9.44 & -9.59 \\ 
        \ding{51} & \ding{51} & -23.49 & -27.41 & -40.79 & -30.56 \\ \hline
    \end{tabular}
\end{table}

 We have demonstrated the training effects of using different loss functions, varying frame lengths, learning rate, and segment type at different stages in the most right figure of Fig.~\ref{fig:Comparisons bdrate}, which is the rate-distortion curve for HEVC Class D from stage8 to stage18. 
 For stage8 to stage11, we use $loss_{recRD}$ as the loss function with input frames from 2 to 6 and increasing input frames significantly improves the performance.  For stage12 to stage15, we use $loss_{all}$ and observe the similar phenomenon. In addition, stage18 shows a significant performance improvement over stage17.

Table~\ref{tab:Ablation} presents the results of the ablation study on the PLS and LSTFFM. The experimental results demonstrate that the incorporation of the PLS leads to a 7.85\% reduction in BD-Rate, while the inclusion of LSTFFM results in a 9.59\% reduction. When both components are utilized together, the BD-Rate reduction reaches 25.03\%. 


\section{conclusion}
\label{sec:majhead}

EV-NVC 
reveals the process of training end-to-end video compression with only maximal six frames and mixed-precision training, and it also introduces PLS 
and LSTFFM to reduce computational complexity and enchance the context model.
Besides EV-NVC will contribute to the advancement of AI flow, which can make AI flow across variable devices.

\vfill\pagebreak

\label{sec:refs}


\bibliographystyle{IEEEbib}
\bibliography{refs}
    
\end{document}